\documentclass[fleqn,twoside]{article}
\usepackage{espcrc2}

\usepackage{graphicx}
\usepackage[figuresright]{rotating}

\newcommand{\AmS}{{\protect\the\textfont2
  A\kern-.1667em\lower.5ex\hbox{M}\kern-.125emS}}

\title{Production and Polarization Effects in Some Tau-Lepton Decays}

\author{A. Ilakovac\address[MCSD]{Physics Department, 
        Faculty of Science, University of Zagreb, pp 331, HR-10002, 
        Zagreb, Croatia}} 
       
\begin{document}

\begin{abstract}
The conditions for the independence of decays of the spin-$1/2$ resonances on the 
production mechanism of the resonances and on polarizations of the incoming and outgoing 
particles are derived and applied in the case of
several tau-lepton decays. 
The necessity for inclusion of the influence 
of the production mechanism in the evaluation of the lepton flavour violating 
decays is stressed.
\vspace{1pc}
\end{abstract}

\maketitle

\section{INTRODUCTION}

The theoretical results for lepton flavour violating (LFV) branching ratios (BR)
are usually given in terms of the decay rates, independently of the
production mechanism of the tau leptons. In this paper the total cross sections
involving  decaying resonances are studied. The cross sections are evaluated exactly 
and assuming the independence of decay processes of the production mechanism.
The deviations of approximative cross section from the exact cross section are analyzed.
The aim of this work is to show that the exact calculation should be applied to LFV processes,
too.

The independence of resonance decay rates of the production 
mechanism has been conjectured in the early sixties \cite{Jac64,PiRo68,Pil79},
and recently it has been refined for a scalar resonance \cite{Lic99}
\begin{eqnarray}
\label{PFLi}
\frac{d\sigma_A}{dM^2d^3P}&=&
\frac{d\sigma_R(M)}{d^3P}\frac{M\Gamma_{R\to A}(M)}{\pi{\cal D}_R(M)}.
\end{eqnarray}
$d\sigma_A$ is the cross section of the process $1+2\to X+(R\to A)$ (1 and 2 are
incoming particles, $R$ is a resonance, $A$ are decay products of $R$,
and $X$ are all particles not included in the $R$ decay).
$d\sigma_R(M)$ is the cross section of the process $1+2\to X+R$ (where $R$ is treated as a
free particle of an unphysical mass $M$).
$\Gamma_{R\to A}(M)$ is the decay rate of $R$
($R$ again treated as a free particle of unphysical mass $M$).
${\cal D}_R(M)$ is the usual denominator of the resonance-$R$ propagator,
\begin{equation}
\label{PropD}
{\cal D}_R(M)\ =\ (M_0^2-M^2)^2+M_0^2\Gamma_R^2(M)
\end{equation} 
($M_0$ is the physical mass of the resonance $R$, $\Gamma_R$ is 
its total decay rate). $P$ is the four-momentum of the resonance $R$ and $M^2=P^2$.
Equation    (\ref{PFLi}) is valid for one decaying resonance in the 
amplitude. It suggests that the process $1+2\to X+(R\to A)$ can 
be viewed as a sequence of two independent processes,
$1+2\to X+R$ and $R\to A$. Therefore, the decay rate of the
resonance $R$ is independent of the production mechanism.
The approximate integral form of Eq. (\ref{PFLi}), valid for 
narrow resonances, which follows from the approximate relation
\begin{equation}
\int_0^\infty M{\cal D}_R(M)^{-2} dM^2=\frac{\pi}{\Gamma_R},
\end{equation} 
is given by
\begin{equation}
\sigma_A\ =\ \sigma_R\frac{\Gamma_{R\to A}}{\Gamma_R}.
\end{equation}
Equation  (\ref{PFLi}) can be generalized for several decaying resonances 
in the amplitude, and this generalization is especially simple for 
narrow resonances.

For a scalar resonance, the product form for the 
total cross section ($\sigma$PF) (\ref{PFLi}) is a consequence of the product form of the 
corresponding matrix element,
\begin{equation}
\label{MSR}
{\cal M}_A\ =\ {\cal M}_R\: i{\cal D}_R^{-1} {\cal M}_{R\to A}, 
\end{equation}
with notation as in (\ref{PFLi}).
${\cal M}_R$ and ${\cal M}_{R\to A}$ are the same matrix elements 
that appear in the evaluation of the cross section $1+2\to R+X$ and $R\to A$.

In the case of spin-$1/2$ and vector resonances or resonances with more
complicated spin-tensor structure, the product form for matrix elements
(\ref{MSR}) is not so simple and the product form for the 
differential cross section 
(\ref{PFLi}) is not generally valid. The product form 
for the differential cross section (\ref{PFLi}) for a vector resonance 
in the amplitude has been 
studied in \cite{Lic99}. 
The conditions for the 
product formula for the differential cross section or total 
cross section for the processes with one 
and two spinor resonances 
are given here.

\section{CONDITIONS FOR $\sigma$PF FOR ONE SPIN- $1/2$ RESONANCE}


For a spin-$1/2$ resonance, the matrix elements ${\cal M}_{R\to A}$, 
${\cal M}_R$ and ${\cal M}_A$ read
\begin{eqnarray}
{\cal M}_{R\to A}&=&\bar{U}_Au_R, \nonumber\\
{\cal M}_R&=&\bar{u}_RU_S, \nonumber\\
{\cal M}_A&=&{\cal D}_R^{-1}\bar{U}_A{\cal P}_RU_S,
\end{eqnarray}
where $u_R$ is the wave function of the resonance, $U_A$ and $U_S$ 
("spinor amplitudes") are
general spinor functions describing the decay products of the the resonance ($A$)
and scattering process $1+2\to X+R$ (without the resonance wave function) and 
${\cal P}_R=\mbox{P\hspace{-.70em}/}+M_0$ is the numerator of the propagator 
of the resonance. Spinor functions $U_A$ and $U_S$ contain all information 
on momenta and polarizations of the decay products $\{A\}$ and particles $1,2,\{X\}$
respectively.
The corresponding decay rates and differential cross sections read
\begin{eqnarray}
\label{GRA}
\Gamma_{R\to A}&=&(16M\pi^2)^{-1}Tr[{\cal P}_RT],\\
\label{sR}
E\frac{d\sigma_R}{d^3P}&=&\frac{\pi}{4F}Tr[{\cal P}_RS],\\
\label{sA}
E\frac{d\sigma_A}{d^3PdM^2}&=&\frac{\pi}{32F}Tr[S{\cal P}_RT{\cal P}_R],
\end{eqnarray}
where $E=(M^2+{\bf P}^2)^\frac{1}{2}$ is the energy of the resonance, 
$F=E_1E_2|{\bf v}_1-{\bf v}_2|$, and
\begin{eqnarray}
\label{T}
T&=&\sum_{\{\lambda_A\}}\int d\mbox{LIPS}_{A}U_A\times \bar{U}_A,\\
\label{S}
S&=&\sum_{\{\lambda_X\}}\int d\phi_X U_S\times \bar{U}_S,
\end{eqnarray}
with ${\{\lambda_A\}}$,  ${\{\lambda_X\}}$ and $d\mbox{LIPS}_{A}$ and 
$d\phi_X$ representing the polarizations
of $\{A\}$ and $\{X\}$ particles, Lorentz invariant
phase space of particles $\{A\}$ and phase space 
of $\{X\}$ particles, respectively.

The most general form of $T$ is 
\begin{equation}
T\ =\ T_1+\mbox{T\hspace{-.60em}/}_2+T_3^{\mu\nu}\sigma_{\mu\nu}+
\mbox{T\hspace{-.60em}/}\gamma_5+T_5\gamma_5.
\end{equation}
The product form for the differential cross section is satisfied if the product of the 
traces from Eqs. (\ref{GRA}) and (\ref{sR}) is equal to the trace from Eq. (\ref{sA}).
The sufficient condition for that equality to be satisfied is 
\begin{equation}
\label{TPF1}
{\cal P}_RT{\cal P}_R\ =\ {\cal P}_R\frac{1}{2}Tr[T{\cal P}_R],
\end{equation}
with the most general solution for $T$,
\begin{equation}
\label{TPF2}
T\ =\ T_1+\mbox{T\hspace{-.60em}/}_2+T_4\bar{\mbox{P\hspace{-.70em}/}}+T_5\gamma_5,
\end{equation}
($\bar{P}=P/M_0$).
Although the condition (\ref{TPF1}) is not the only one 
that realizes the product form for the differential cross section (\ref{TPF1})
for one resonance in the scattering process $1+2\to R+X$,
it is the only one 
that remains for more than one decaying resonance in the scattering 
process.

If the summation over all polarizations of $\{A\}$ particles and
integration over complete $d\mbox{LIPS}_{A}$ phase space is performed,
$T$ depends only on the momentum of the resonance $P$, and its most general
form,
\begin{equation}
\label{T4piSp}
T\ =\ T_1+T_2\bar{\mbox{P\hspace{-.70em}/}}+T_4\bar{\mbox{P\hspace{-.70em}/}}+T_5\gamma_5
\end{equation}
(form factors $T_i$ depend only on $P^2$) satisfies the product form condition
(\ref{TPF1}). Therefore, if an experiment is performed with a $4\pi$ detector,
and polarizations are not measured, the $\sigma$PF is satisfied for processes 
with one decaying spin-$1/2$ resonance. The result can be generalized for processes
with any number of decaying spin-$1/2$ resonances. 

In the following, three possible situations which deviate from the situation 
described by Eq. (\ref{T4piSp}), will be discussed,\\
1. polarization of the decay products $\{A\}$ of the resonance,\\
2. a process with two or more spin-$1/2$ resonances,\\ 
3. partial integration over phase space for processes
with two spin-$1/2$ resonances.\\

\section{$\sigma$PF AND POLARIZATION}

The description of the polarization of massless spin-$1/2$ particles, massive spin-$1/2$ and 
vector particles and massless vector particles is different. Helicity of the 
massless spin-$1/2$ particles is completely determined by the $V\pm A$ vertices and 
no additional four-vector is needed for the description of the particle polarization.
For massive spin-$1/2$ and vector particles one has to introduce additional four-vector
along with the particle four-momentum for a complete description of the particle state. 
For instance, for a massive spin-$1/2$ particle of mass $m$ and four-momentum $p$, the four 
vectors describing the particle state are,
\begin{equation}
\label{FPol4v}
p\ =\ (E,p{\hat{\bf p}}), s\ =\ (p,E{\hat{\bf p}})/m.
\end{equation}
(${\hat{\bf p}}$ is the unit vector in the direction of ${\bf p}$). The direction 
of the spacial part  of the four-vectors $p$ and $s$ are the same. That follows from 
the definition of the helicity quantization axis and the fact that the little group of the
massive particle is attached to its rest frame. For massless vector particles,
the additional four-vector that is necessary for the description of a helicity of the state,
$\eta$, is independent of the four-momentum of the particle, and it represents the 
laboratory frame in which the quantization of the massless-vector-particle spin is performed. 
In the laboratory frame it is a pure time-like four-vector.

If among the polarized particles of the decay products 
of the resonance $R$ only spin-$1/2$ and massive 
vector particles appear, the complete integration 
over $d\mbox{LIPS}_A$ gives $T$ again the structure (\ref{T4piSp}) that assures 
$\sigma$PF. The presence of polarized massless vector particles among the decay 
products of the resonance $R$ leads to additional terms of the $T$ even after the 
complete integration over $d\mbox{LIPS}_A$,
\begin{eqnarray}
\label{PFviol}
T& =& T_1+T_{2A}\bar{\mbox{P\hspace{-.70em}/}}+T_{2B}\mbox{$\eta$\hspace{-.45em}/}+
T_3\sigma_{\mu\nu}\eta^\mu \bar{P}^\nu
\nonumber\\
&+&T_{4A}\bar{\mbox{P\hspace{-.70em}/}}+
T_{4B}\mbox{$\eta$\hspace{-.45em}/}\gamma_5+T_5\gamma_5,
\end{eqnarray}
which cannot be removed by the most general gauge transformation of polarization 
vectors of the type,
$\varepsilon(p,\lambda_i)\to \varepsilon(p,\lambda_i)+\alpha_i p$,
where $p$ is the four-momentum of the massless vector particle and $\alpha_i$ are constants.
The fourth and sixth term in Eq. (\ref{PFviol}) violate the product form condition
(\ref{TPF1}).

\section{$\sigma$PF FOR TWO RESONANCE 
PROCESSES} 

The expected $\sigma$PF equation for a two narrow-resonance process is 
\begin{eqnarray}
\label{PF2R}
\sigma_{AB}&=&
\sigma_{R_1R_2}\frac{\Gamma_{R_1\to A}}{\Gamma_{R_1}}
\frac{\Gamma_{R_2\to B}}{\Gamma_{R_2}}
\ \equiv \sigma_{AB}^{PF}
\end{eqnarray}
($\sigma_{AB}\equiv\sigma_{1+2\to(R_1\to A)(R_2\to B)}$,
$\sigma_{R_1R_2}\equiv\sigma_{1+2\to R_1R_2}$).
The relevant matrix elements for studying $\sigma$PF are
\begin{eqnarray}
{\cal M}_A&\equiv&{\cal M}_{R_1\to A}\ =\ \bar{U}^Au^{R_1},\\
{\cal M}_B&\equiv&{\cal M}_{R_2\to B}\ =\ \bar{u}^{R_2}U^B,\\
{\cal M}_{R_1R_2}&\equiv&{\cal M}_{1+2\to R_1R_2}\ =\ \bar{u}^{R_1} s {u}^{R_2},\\
{\cal M}_{AB}&=&\bar{U}^A{\cal P}^{R_1} s {\cal P}^{R_2} U^B
{\cal D}_{R_1}^{-1} {\cal D}_{R_2}^{-1}
\end{eqnarray}
($u^{R_1}$ and $u^{R_2}$ are the wave functions of the resonances,
$U^A$ and $U^B$ describe the resonance products,
${\cal P}^{R_1}$ and ${\cal P}^{R_2}$ are the numerators of the propagators of the resonances,
${\cal D}_{R_1}$ and ${\cal D}_{R_2}$ are the corresponding denominators of the propagators,
and $s$ is $1+2\to R_1R_2$ amplitude without the resonance wave functions).
Squaring the amplitude, using the usual trick for the subdivision of the phase space
of outgoing particles (\cite{Lic99,BK73}), 
\begin{eqnarray}
dLIPS_{AB}& =& \frac{1}{(2\pi)^2}dM^2_{R_1}dM^2_{R_2}
dLIPS_{R_1R_2}
\nonumber\\
&&dLIPS_AdLIPS_B,
\end{eqnarray}
integrating out the delta functions, and 
integrating over the squares of unphysical resonance masses, 
one obtains the 
approximative expression (valid for narrow resonances) 
for the total cross section (LHS of Eq. (\ref{PF2R}))
\begin{eqnarray}
\label{sigAB}
\lefteqn{\sigma_{AB}\ \approx\ \frac{P\lambda^{-\frac{1}{2}}F^{-1}}
{2^8\pi^2\Gamma_{R_1}\Gamma_{R_2}M^0_{R_1}M^0_{R_2}}
\int d\Omega_{R_1}dL_AdL_B}
\nonumber\\
&&\mbox{\hspace*{-1.5em}}\sum_{ij}S_{ij}^{IJ}
Tr[{\cal P}^{R_1}t^A{\cal P}^{R_1}\tilde{\gamma}_{iI}
 {\cal P}^{R_2}t^B{\cal P}^{R_2}\tilde{\gamma}_{jJ}]\\
&=& \frac{P\lambda^{-\frac{1}{2}}F^{-1}}
{2^6\pi^2\Gamma_{R_1}\Gamma_{R_2}}
\int d\Omega_{R_1}
\nonumber\\
&&{\hspace*{-1.5em}}\sum_{ij}S_{ij}^{IJ}
Tr[{\cal P}^{R_1}\tilde{\Gamma}_A{\cal P}^{R_1}\tilde{\gamma}_{iI}
 {\cal P}^{R_2}\tilde{\Gamma}^B{\cal P}^{R_2}\tilde{\gamma}_{jJ}]
\end{eqnarray}
($\lambda=(p_1+p_2)^2$, $P=|{\bf P}_{R_1}|=|{\bf P}_{R_2}|$,
$M^0_{R_1}$ and $M^0_{R_2}$ are physical masses of the resonances;
$d\Omega_{R_1}$ is differential solid angle of  $\hat{{\bf P}}_{R_1}$, 
$dL_A$ and $dL_B$ are remnants of $d\mbox{LIPS}_{A,B}$ phase
spaces after integration over delta functions; 
${\cal P}^{R_a}=\mbox{P\hspace{-.70em}/}_{R_a}\pm M_{R_a}$, $a=1,2$,
are numerators of $R_a$ propagators -- plus sign is for a particle resonance and
minus sign for an antiparticle resonance;
\begin{eqnarray}
t^A&=&\gamma_0U_A^\dagger\times U_A,\nonumber\\
t^B&=&\gamma_0U_B^\dagger\times U_B,\nonumber
\end{eqnarray}
are absolute squares of the spinor amplitude decay products of resonances $R_{1,2}$,
\begin{eqnarray}               
\sum_{ij}S_{ij}^{IJ}\tilde{\gamma}_{iI}\tilde{\gamma}_{jJ}&\equiv&
S\ =\ \gamma_0 s^\dagger\gamma_0\times s,
\end{eqnarray}
$\tilde{\gamma}_{i}$ are Dirac algebra matrices and $I$ is a set of 
Lorentz indices assigned to a particular $\tilde{\gamma}_{i}$ matrix; 
$\sum_{ij}S_{ij}^{IJ}$ are tensor functions which are part of the
absolute square of the spinor amplitude $s$; the quantities
\begin{equation}
\tilde{\Gamma}_{A,B}\ =\ \int dL_{A,B}t^{A,B}
\end{equation} 
are "spinor decay rates" of resonances $R_{1,2}$ because
$\Gamma_{{R_{1,2}}\to A,B}=\frac{1}{2M_{R_{1,2}}}Tr[{\cal P}^{R_1,R_2}\tilde{\Gamma}_{A,B}]$ 
). 
The expression for the RHS of Eq. (\ref{PF2R}) is obtained from 
(\ref{sigAB}) by replacement
\begin{eqnarray}
\label{sigABPF}
\lefteqn{\sum_{ij}S_{ij}^{IJ}
Tr[{\cal P}^{R_1}t^A{\cal P}^{R_1}\tilde{\gamma}_{iI}
 {\cal P}^{R_2}t^B{\cal P}^{R_2}\tilde{\gamma}_{jJ}]\ \to}
\nonumber\\&&
\sum_{ij}S_{ij}^{IJ}
Tr[{\cal P}^{R_1}\tilde{\gamma}_{iI}{\cal P}^{R_2}\tilde{\gamma}_{jJ}]
\nonumber\\&&
\times Tr[t^A{\cal P}^{R_1}]Tr[t^B{\cal P}^{R_2}].
\end{eqnarray}
The $\sigma$PF equation leads to the equality of LHS and RHS of Eq. (\ref{sigABPF}).
It can be shown that the only way to satisfy that equality is that the
following two conditions are satisfied,
\begin{eqnarray}
\label{TPFC2R1}
({\cal P}^{R_1}T^A{\cal P}^{R_1})
&=& {\cal P}^{R_1}\frac{1}{2}Tr(T^A{\cal P}^{R_1}),\nonumber\\
({\cal P}^{R_2}T^B{\cal P}^{R_2})
&=& {\cal P}^{R_2}\frac{1}{2}Tr(T^B{\cal P}^{R_2})
\end{eqnarray}
($T^{A,B}=t^{A,B}$ or $T^{A,B}=\int dL_{A,B}t^{A,B}$). Eqs. (\ref{TPFC2R1}) are
identical in form to Eq. (\ref{TPF1}), and their most general 
solutions have the same form as
(\ref{TPF2}),
\begin{eqnarray}
\label{TPFC2R2}
T^A&=&T_1^A+\mbox{T\hspace{-.60em}/}_2^A+
T^A_4\mbox{P\hspace{-.70em}/}^A_4\gamma_5+T_5^A\gamma_5\nonumber,\\
T^B&=&T_1^B+\mbox{T\hspace{-.60em}/}_2^B+
T^B_4\mbox{P\hspace{-.70em}/}_4^B\gamma_5\!+\!T_5^B\gamma_5.
\end{eqnarray}
The generalization of $\sigma$PF conditions for more than 
two spin-$1/2$ resonances is obvious.

\section{ANALYSIS OF $\sigma$PF for $e^-e^+\to(\tau^-\to A)(\tau^+\to B)$
PROCESSES}

From the above analysis follows that to prove the some process
containing only spin-$1/2$ resonances satisfies
$\sigma$PF one has to show that 
all absolute squares of the spinor amplitudes of spin-$1/2$ resonances included 
in the process, or any phase space integral (partial or complete)
of these quantities must satisfy the condition (\ref{TPF1}).
That rule will be used to study several $e^-e^+\to(\tau^-\to A)(\tau^-\to B)$
ptocesses with respect to the condition (\ref{TPF1}).
Three processes are chosen for the analysis,
$e^-e^+\to(\tau^-\to\pi^-\nu_\tau)(\tau^+\to\pi^+\bar{\nu}_\tau)$,
$e^-e^+\to(\tau^-\to l^-\bar{\nu}_l\nu_\tau)(\tau^+\to l^+\nu_l\bar{\nu}_\tau)$
and
$e^-e^+\to(\tau^-\to\pi^-\nu_\tau)(\tau^+\to l^+\gamma)$.
First two are Standard Model processes, while the third
belongs to the physics beyond the Standard Model. The first one contains one
observable scalar per decaying resonance, the second contains one observable
lepton per resonance, and the third comprises one resonance as 
in the first case while the other resonance decays into observable lepton
and photon. The parts of the phase space of the unobservable particles 
can be integrated. In the following analysis  simple 
model of LFV -- Standard model with additional heavy neutrinos 
{\cite{LFV}) is used. 

The spinor decay rates
involved
in the processes read
(complete phase space integration and summation over all polarizations is
assumed):
\begin{eqnarray}
\tilde{\Gamma}_{\tau\to\pi\nu_\tau}\!\!\! & =&\!\!
\bar{G}_F^2\bar{f}_\pi^2\frac{|V_{ud}|^2}{2^3\pi}
\Big(1-\frac{m_\pi^2}{m_\tau^2}\Big)
\bar{\mbox{P\hspace{-.70em}/}}_\tau P_L,
\nonumber\\
\tilde{\Gamma}_{\tau\to l^-\bar{\nu}_l\nu_\tau}\!\!\! & =&\!\!
\frac{\bar{G}_F^2}{3\times 2^6\pi^3}\bar{\mbox{P\hspace{-.70em}/}}_\tau P_L, 
\nonumber\\
\tilde{\Gamma}_{\tau\to l^-\gamma}\!\!\! & =&\!\!
\frac{\alpha_W^3s_W^2}{2^8\pi^2\bar{M}_W^4}(1-\bar{m}_l^2)^3
\nonumber\\
&&\bar{\mbox{P\hspace{-.70em}/}}_\tau(P_R+\bar{m}_l^2P_L)F_{LFV}^2.
\end{eqnarray}
Barred quantities are dimensionless. They are obtained from the corresponding 
dimensionfull quantities multiplying them by a power of the tau mass. $G_F$ is
Fermi constant, $\alpha_W=g^2/(4\pi)$; $F_{LFV}=(\sum_i B_{\tau i}^*B_{li}$
$G(\lambda_i))^2$ is a factor which contains information on the magnitude of 
LFV -- see Refs. \cite{LFV}.
Obviously, all spinor decay rates satisfy the $\sigma$PF condition 
(\ref{TPF1}). Notice the similarity in the structure of the spin dependent 
parts of all three $\tilde{\Gamma}$-s.

The approximative expressions 
(the numerator of the $W$-boson propagator s replaced by
$M_W^2$) for the absolute squares of spinor amplitudes for these three processes
read 
\begin{eqnarray}
t_{\tau\to \nu_\tau\pi}\!\!\!&\approx&\!\!
m_\tau\frac{g^4}{8}\frac{\bar{f}_\pi^2}{\bar{M}_W^4}V_{ud}^2
(\bar{\mbox{p\hspace{-.50em}/}}_\pi-\bar{m}_\pi^2\bar{\mbox{P\hspace{-.70em}/}}_\tau)
\nonumber\\
t_{\tau\to \nu_\tau\nu_ll}\!\!\!&\approx&\!\!
m_\tau^{-1}g^4\bar{M}_W^{-4}(1-\bar{s}_{\tau l}-\bar{s}_{\nu_ll})
\bar{\mbox{p\hspace{-.50em}/}}_{\nu_l}P_L
\nonumber\\
t_{\tau\to l\gamma}\!\!\!&=&\!\!
m_\tau\frac{s_W^2(1-\bar{m}^2_l}{2^{9}\pi^4\bar{M}^4_W}F_{LFV}^2
\nonumber\\
\!\!\!&&\!\!           
\bar{\mbox{p\hspace{-.50em}/}}_\gamma(P_L+\bar{m}_l^2P_R)
\end{eqnarray}
($\bar{s}_{\tau l}=(\bar{p}_\tau-\bar{p}_l)^2$, 
$\bar{s}_{\nu_ll}=(\bar{p}_\nu+\bar{p}_l)^2$),
The structures of the spin dependent parts of all three $t$-s are similar.
Further, none  of three absolute squares of spinor amplitudes satisfies the
$\sigma$PF condition (\ref{TPF1}). 
Therefore, a deviation from the $\sigma$PF is expected if the 
complete integration over phase space is not peformed. The phase space
measures of all
three processes contain the differentials of solid angles of the two observable particles.
For the process 
$e^-e^+\to(\tau^-\to l^-\bar{\nu}_l\nu_\tau)(\tau^+\to l^+\nu_l\bar{\nu}_\tau)$,
we performed the following integrations over phase space. 
We defined the difference of the differentials of the RHS and LHS of 
Eq. (\ref{PF2R})
\begin{equation}
d\Delta\sigma\ =\ d\sigma_{A,B}-\frac{1}{\Gamma_\tau^2}d\sigma_{R_1R_2}
d\Gamma_Ad\Gamma_B.
\end{equation}
Next, the unit spheres of the 
solid angle variables of the measurable particles were subdivided into two hemispheres 
(left $L$ and right $R$ with respect  to the
direction of the incoming electron). So  four sectors with 
respect to the solid angle integrations of observable particles were obtained ($LL$, $LR$,
$RL$ and $RR$). Further, we performed a complete integration
over one of the four parts of the phase space and at the same 
time over the remaining part of the phase space.
The results evaluated for center of mass energy $\lambda^{1/2}=\sqrt{20}m_\tau$ 
are given in the following table.\\[.2cm]

\noindent
Table 1.\\[.1cm]
{\small The deviation of the cross section from the\\
 product form}\\[-.1cm]
\begin{tabular}{lr}\\ 
\hline
sector        & $\Delta\sigma(GeV^{-2})\ $\\ 
\hline
$LL$          & $-7.67\times 10^{-10}$\\ 
$LR$          & $7.44\times 10^{-10}$\\
$RL$          & $7.82\times 10^{-10}$\\
$RR$          & $-7.47\times 10^{-10}$\\
$LL+LR+RL+RR$ & $0.11\times 10^{-10}$\\
\hline
\end{tabular}\\[.2cm]

\noindent
The integration is ten dimensional.
The integration was performed by a simple, non-adaptive Monte-Carlo 
program. Therefore, the result obtained for the sum of the four contributions 
can be considered as consistent with zero (it tends to zero as one enlarges the
number of integration points).
On the other side, the cross sections for the total decay rate 
for $e^-e^+\to(\tau^-\to l^-\bar{\nu}_l\nu_\tau)(\tau^+\to l^+\nu_l\bar{\nu}_\tau)$,
evaluated using the 
product form formula (LHS of Eq. (\ref{PF2R})), is 
\begin{equation}
\sigma^{PF}\ =\ 4.42\times 10^{-9} GeV.
\end{equation}

The relative deviation of from the $\sigma$PF is of the order of $20\%$.
The similarity of the structures of the spin dependent parts of all three $t$-s
indicates that the relative deviation of from the $\sigma$PF for other 
two processes is of the same order of magnitude.

In the literature, LFV decay rates are usually theoretically evaluated assuming the
$\sigma$PF. On the other side, measurements of LFV processes are performed by
detectors which are not completely $4\pi$
detectors. Therefore, in principle,  
one should include the dependence of the LFV decay rates on the 
production mechanism when presenting the theoretical results.

\section{CONCLUSION}
The conditions for the product form of a decay rate ($\sigma$PF) of an 
amplitude containing one or more spin-$1/2$ resonances have
been found. The analysis of the $\sigma$PF has been performed for 
one and two resonances, for 
unpolarized particles and complete integration over phase space of 
outgoing particles, for polarized particles and complete integration over
phase space of outgoing particles, and for unpolarized particles 
and partial integration over 
phase space. It has been found that the $\sigma$PF is not 
satisfied if 
decay products of the resonance contain polarized 
massless spin-$1$ particle(s) even for complete integration over 
phase space or if partial integration over phase space is performed,
regardless of the polarization of the particles. 
The formalism has been applied to three $e^-e^+\to(\tau^-\to A)(\tau^+\to B)$
processes, two Standard Model processes and one lepton flavour violating (LFV)
process. These three examples show explicitely that complete 
integration over phase space and summation over all polarizations
leads to the $\sigma$PF, but if partial integration over phase space
is performed, large deviations from the $\sigma$PF are obtained.
A comment on a possible improvement of the theoretical results for LFV decay 
rates, including their dependence on the
production mechanism, is given.\\

\noindent
{\bf Acknowledgements} The author is grateful to the Workshop organizers and
particularly to A. Seiden for the opportunity to attend the Workshop and
to present this talk. Author is indebted to S. Eidelman for suggesting the
study the problem presented here, and also thanks Z. Was,
S. Eidelman, A. Czarnecki, K. Inami, A. Stahl and M. Hayashii for very instructive 
discussions.\\

\noindent
{\bf Note} After the author sent this paper to arXiv.org, Y. Okada draw his attention to the 
paper \cite{KO01}, in which spin correlations in the leptonic LFV tau-lepton decays were 
discussed in detail.\\


\begin{thebibliography}{9}
\bibitem{Jac64} J.D. Jackson, Nuovo Cimento 34 (1964) 1544.
\bibitem{PiRo68} J. Pi\v s\'ut, M.Ross, Nucl. Phys. B 6 (1968) 325.
\bibitem{Pil79} H.M. Pilkuhn,  Relativistic Particle Physics, 
 Springer-Verlag New York Inc., 1979.
\bibitem{Lic99} P. Lichard, Acta Phys. Slov 49 (1999) 215.
\bibitem{BK73} E. Byckling and K. Kajantie, Particle Kinematics,
John Wiley \& Sons, London, 1973.
\bibitem{LFV} R.N. Mohapatra and J.W.F. Valle, Phys. Rev. D 34 (1986) 1642;
Bernabeu et al. Phys. Lett B 187 (1987) 303; 
A. Pilaftsis, Z. Phys. C 55, 275 (1992), 
A. Ilakovac and A. Pilaftsis, Nucl. Phys. B 437 (1995) 491;
 A. Ilakovac, Phys. Rev. D {\bf 62} (2000) 36010. 
\bibitem{KO01} R. Kitano and Y. Okada, Phys. Rev. D 63 (2001) 113003.
\end{thebibliography}
\end{document}